\documentclass[12pt,preprint]{aastex}

\shorttitle{}
\shortauthors{}

\begin{document}

\newcommand{\php}[0]{\phantom{--}}
\newcommand{\kms}[0]{km~s$^{-1}$}

\title{THE ORIGIN OF RADIO SCINTILLATION IN THE LOCAL INTERSTELLAR MEDIUM}


\author{Jeffrey L. Linsky\altaffilmark{1}\altaffiltext{1}{JILA,
University of Colorado and NIST, Boulder, CO 80309-0440},
Barney J. Rickett\altaffilmark{2}\altaffiltext{2}{Department of
Electrical and Computer Engineering, University of California San
Diego, La Jolla, CA 92093-0407}, and
Seth Redfield\altaffilmark{3,4}\altaffiltext{3}{Department of
Astronomy and McDonald Observatory, University of Texas, Austin, TX
78712-0259}\altaffiltext{4}{Hubble Fellow}}
\email{jlinsky@jila.colorado.edu, bjrickett@ucsd.edu, 
sredfield@astro.as.utexas.edu}

\begin{abstract}

We study three quasar radio sources (B1257-326, B1519-273, and J1819+385)
that show large amplitude intraday
and annual scintillation variability produced by the Earth's motion
relative to turbulent-scattering screens located within a few parsecs
of the Sun.  We find that the lines of sight to these sources pass
through the edges of partially ionized warm interstellar clouds where
two or more clouds may interact. From the gas flow vectors of these
clouds, we find that the relative radial and transverse velocities of
these clouds are large and could generate the turbulence that is
responsible for the observed scintillation.  For all three sight lines
the flow velocities of nearby warm local interstellar clouds are
consistent with the fits to the transverse flows of the radio
scintillation signals.

\end{abstract}


\section{OBSERVATIONS OF FAST RADIO SCINTILLATION}

Interstellar scintillation (ISS) is the apparent variation in flux
density of very compact radio sources due to propagation through the
irregular refractive index of the ionized interstellar medium.  The
phenomenon was discovered and investigated through its 100\%
variations in pulsar signals seen on time scales from minutes to days.
Most extragalactic radio sources have angular diameters much too large
to scintillate, in the same way that the angular size of planets
suppresses their atmospheric twinkling.  However, the most compact
parts of some active galactic nuclei do fluctuate at a level of
1--10\% over times shorter than a few days in a phenomenon called
intraday variation (IDV) \citep[see,][]{heeschen87,quirrenbach92}.
Although this was initially thought to be intrinsic variation of the
sources interpreted as due to emission from regions with brightness
temperatures very much higher than the inverse Compton limit, it was
subsequently explained as ISS somewhere within the 1~kpc thick layer
of the ``warm ionized medium'' (WIM) of our Galaxy, and implied
maximum source brightness temperatures near the inverse Compton limit
\citep[e.g.,][]{rickett07}.

However, among intraday variables, a few sources stand out as having
unusually fast high-amplitude variations at frequencies of 3--8~GHz.
With time scales less than a few hours, they have been called
``intrahour variables'' (IHV), the best studied of which are
B1257--326, B1519--273, and J1819+385.  Their short time scales can
only be explained by scintillation in a local region whose electron
density has much greater root-mean-squared variation than in the
general WIM.  The distances to the scattering are estimated to be in
the range 1--30~pc.  However, until now no specific site in the local
ISM has been identified as responsible.  A common feature of the IHV
observations is that there is a systematic annual variation in their
time scales due to a changing velocity of the Earth relative to the
interstellar-scattering medium.  Indeed several observers have
estimated the characteristic transverse velocities of the medium
relative to the Sun needed to fit the ISS observations
\citep{Bignall2006,Dennett2003,Jauncey2003}.

It is also interesting that only a very few sources varied on hourly 
time scales in the 5~GHz VLA survey of 525 compact flat-spectrum AGNs,
even though about half of them varied at a low level on times of
several days \citep{Lovell2007}.  Thus the regions responsible for IHV
cover a very small fraction of the sky.   The basic goal of this paper
is to compare the velocities from the rapid ISS sources 
with the velocities of the warm
local interstellar clouds identified by \citet{Redfield2007} 
and so find the regions of the local ISM responsible for the IHV.
We find that the turbulence responsible for the scintillation
may be generated by cloud-cloud interactions that could also 
enhance the ionization at cloud interfaces.

\section{WARM CLOUDS IN THE LOCAL INTERSTELLAR MEDIUM}

The Sun is located in a $\sim$100\,pc radius volume devoid of dense
interstellar material known as the Local Bubble \citep{lallement03}.
Within the Local Bubble, a group of warm partially ionized
clouds is found in the immediate ($\lesssim$20\,pc) vicinity of the
Sun and is often referred to as the Local Interstellar Medium (LISM).  This
collection of gas is comprosed of distinct components, or clouds.
This is demonstrated by the fact that LISM sight lines often exhibit
multiple distinct absorbers, on average $\sim$1.7 absorbers per sight
line \citep{redfield2004}; therefore any given line of sight typically
has between 1 and 3 components.  In addition, physical characteristics
(e.g., temperature, metal depletion, etc.) and cloud kinematics are
spatially correlated, that is, individual clouds have unique
properties and dynamics.

\citet{Redfield2007} used the largest LISM observational database,
based on high spectral resolution ultraviolet and optical absorption
measurements, to investigate the kinematical properties of the closest
LISM clouds.  They derived three-dimensional velocity vectors for 15
clouds, which all reside within 15\,pc, including the Local
Interstellar Cloud (LIC) and Galactic (G) Cloud, the two closest
clouds, whose dynamics were first calculated by \citet{lallement92}
and \citet{lallement95}.  The 15 velocity vectors are roughly
parallel, indicating that they share a common origin or driver, an
issue discussed by \citet{frisch02}.  However, the velocity vectors
have a range of velocities (0--50 km~s$^{-1}$ relative to the Local
Standard of Rest), which implies that cloud-cloud collisions and
interactions will play a prominent role in the physical properties of
the LISM.

The outer edges of LISM clouds are likely sites for interesting
phenomena such as scintillation screens.  It is in these regions that
the kinematic differences between neighboring clouds are most extreme
and the effects of any interaction most prominent.  For example,
shearing flows between clouds with just moderately different velocity
vectors may induce turbulence in the interaction zone between the
clouds.  In addition, the edges of the clouds should have higher
ionization levels, as compared to the bulk of the cloud, due to the
lack of significant shielding from ionizing radiation sources
such as the B-star $\epsilon$ CMa and the hot white dwarfs Feige 24,
HZ 43, and G191-B2B \citep{vallerga98}.  These regions are prime
locations for enhanced electron density and high levels of turbulence
and are therefore ideal sites for local scintillation screens.

\begin{figure}
\plotone{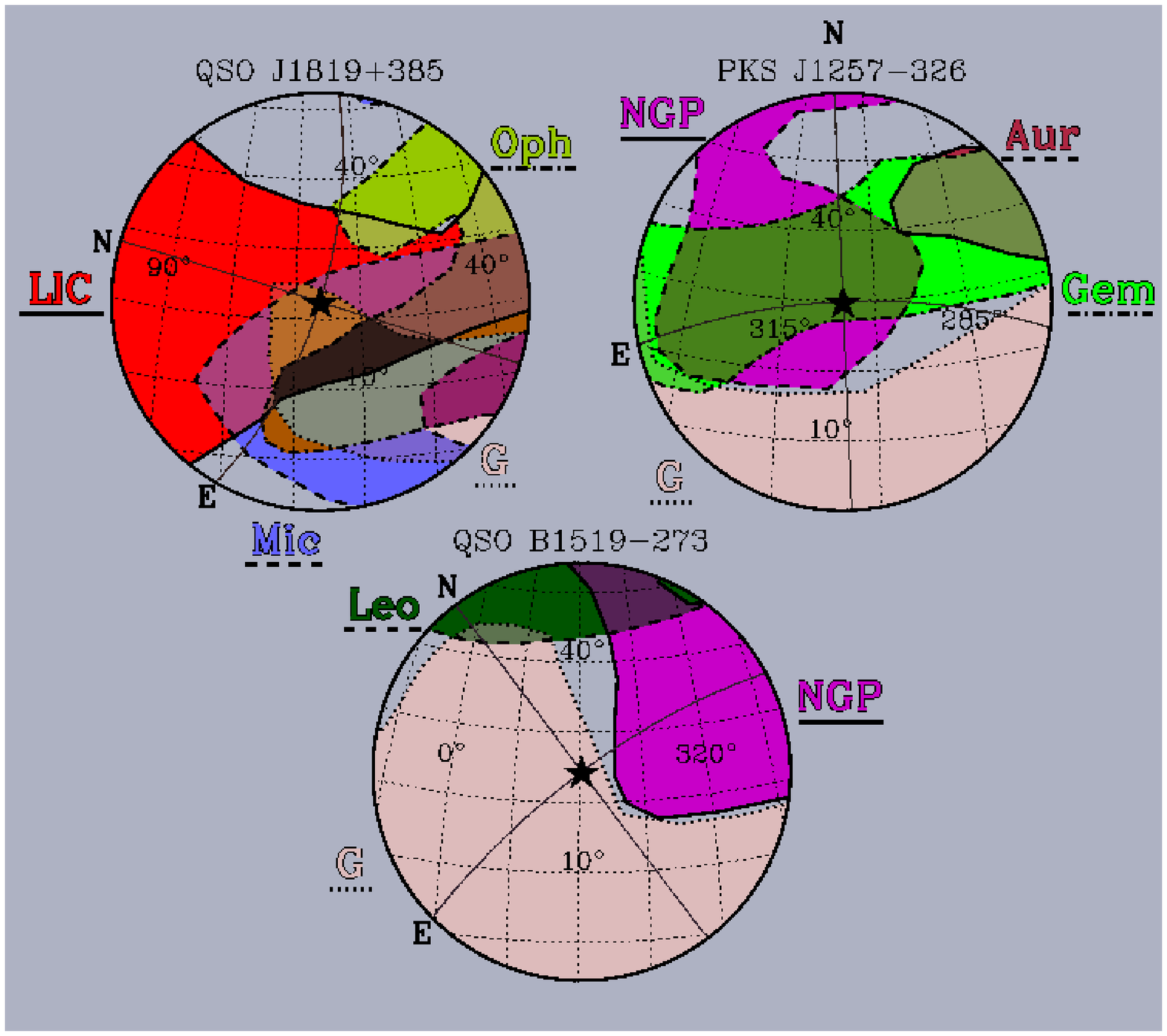}
\caption{The morphology, in Galactic coordinates, of local
interstellar clouds located within $30^{\circ}$ of the three quasars
showing large amplitude intraday and annular scintillation
variability. The interstellar clouds identified by
\citet{Redfield2007} are color coded and labeled.  Celestial North
and East are identified in each plot.  \label{fig:fig1}}
\end{figure}

The distribution of the 15 dynamical clouds in Galactic coordinates is
shown in Figure 19 of \citet{Redfield2007}. We plot in
Figure~\ref{fig:fig1} $30^{\circ}$ circular regions centered on the
three quasars, showing the boundaries of nearby clouds that could be
the sites of scattering screens. In Figure~\ref{fig:fig2}, regions
where several clouds lie along the line of sight are colored, and the
shading is related to the magnitude of the maximum velocity difference
between any pair of clouds along that particular sight line.  Since we
know the three-dimensional velocity vector for all 15 clouds, we can
decompose the velocity into transverse ($v_{\rm l}$ and $v_{\rm b}$)
and radial ($v_{\rm r}$) components.  The differential velocity shown
in Figure~\ref{fig:fig2} is calculated from the differences between
two clouds along a line of sight, $\sqrt{(\Delta v_{\rm r})^2 +
(\Delta v_{\rm l})^2 + (\Delta v_{\rm b})^2}$.  All three
scintillation sight lines discussed in this work are also shown in
Figure~\ref{fig:fig2}, along with two additional known 
but less well studied intrahour
variables.  Note that all are located in regions where multiple clouds
are found and the probability of cloud interactions is high.  The
differential velocities in these regions are also significant ($>$10
km~s$^{-1}$ and even as high as $\sim$45 km~s$^{-1}$).  The location
of the scintillation sight lines is quite suggestive, but a more
stringent test of the connection between scintillation screens and the
warm LISM clouds can be achieved using the recently determined
velocity vectors of the LISM clouds.  The transverse velocity can now
be calculated and directly compared to the transverse flow measured
toward local scintillation screens. Without a velocity vector, the
only observable velocity is the radial component.  The velocity
components of LISM clouds located along or near the three
scintillation sight lines studied in this work are listed in
Table~\ref{tab:lismvecs}.  The comparison of scintillation transverse
velocities with the transverse velocities of nearby warm LISM clouds
is the subject of the following sections.

\section{FITTING TO THE SCINTILLATION TIME SCALE\label{sec:fits}}

Observers have characterized the scintillation in flux density by
forming its auto-correlation function and defining a time scale as the
time lag where the correlation falls to half of its (noise-corrected)
value at zero lag.  When this time scale is plotted against day of
year, an annual variation can be seen for the IHV sources, see
Figures~\ref{fig:1257}--\ref{fig:1819}.  The IHV itself and the annual
variation of its time scale are produced by the Earth's motion
relative to the scintillation pattern of focused and defocused waves
produced by the turbulent plasma.  The time scale for interstellar
scintillation, $t_{\rm iss}$, is determined by the characteristic
spatial scale of the ISS diffraction pattern divided by the \it
transverse \rm velocity of the Earth relative to the pattern. Two
processes cause an annual variation in $t_{\rm iss}$ -- the changing
velocity of the Earth relative to the interstellar plasma and the
possibility that the diffraction pattern is not statistically circular
but is characterized by an ellipse with axial ratio $A$, geometric
mean spatial scale $s_{\rm iss}$, and orientation angle of the major
axis $\theta_{A}$ with respect to East (through North).  The
noncircular diffraction pattern could be produced either by an
anisotropic source structure or by anisotropic scattering.  Thus both
the effective pattern scale and velocity vary as the Earth orbits the
Sun.

The algebraic details are as follows.  If $\vec{V}_{\rm E}$ is the
velocity of the Earth relative to the Sun on a given day and
$\vec{V}_{\rm ism}$ is the (unknown) velocity of the interstellar
plasma relative to the Sun, we write RA and DEC components of the
Earth's transverse velocity relative to the interstellar plasma (and
to the scintillation pattern) as:
\begin{equation}
V_{{\rm iss},\alpha} = V_{{\rm E},\alpha} - V_{{\rm ism},\alpha} \; ;
V_{{\rm iss},\delta} = V_{{\rm E},\delta} - V_{{\rm ism},\delta}.
\end{equation}
Taking the spatial correlation function of the
intensity scintillation to be a simple function of a quadratic form in 
orthogonal
transverse coordinates, the resulting ISS time scale is:
\begin{equation}
t_{\rm iss} = s_{\rm iss}/ (a V_{{\rm iss},\alpha}^2 + 
b V_{{\rm iss},\delta}^2 + 
c V_{{\rm iss},\alpha} V_{{\rm iss},\delta})^{0.5} ,
\label{eq:tiss}
\end{equation}
where for the ellipse specified as above, the coefficients 
of the quadratic form are
$a = {\cos}^2 \theta_{A}/A + A {\sin}^2 \theta_{A}$, 
$b = {\sin}^2 \theta_{A}/A + A {\cos}^2 \theta_{A}$, 
$c = 2{\sin}\theta_{A} {\cos}\theta_{A}(1/A - A)$, and
$s_{\rm iss}$ is the geometric mean spatial scale on which the 
intensity correlation falls to 1/e.
Equation (\ref{eq:tiss}) is mathematically the same 
as equation (3) of \citet{Bignall2006}.

The radio observations are typically a set of $t_{\rm iss}$ versus 
date measurements, as shown by the points in the left hand panels of Figures
\ref{fig:1257}, \ref{fig:1519} and \ref{fig:1819}.
The fitting task is to find 
the three ISS parameters ($A$, $\theta_{A}$, $s_{\rm iss}$) and the
two plasma velocities ($V_{{\rm ism},\alpha},V_{{\rm ism},\delta}$)
that give the best fit using this model equation.  It can be shown that there
are five independent coefficients in the general case, but that
the five physical parameters depend nonlinearly on the $t_{\rm iss}$ data 
and so cannot be
estimated independently.  Thus, while we fitted for the best of these
five parameters, we also investigated the error surface by stepping through 
a grid of values for ($V_{{\rm ism},\alpha},V_{{\rm ism},\delta}$)
and at each step fitting for the remaining three ISS parameters;
we constrained the axial ratio to be less than 10, since the fit becomes 
degenerate for very large axial ratios, in a fashion similar to that used 
by \citet{Dennett2003}.  The right
hand panels of the same figures show that for all three sources
there is a satisfactory fit in an extended valley in ISM velocity space.

\subsection{\it Quasar B1257$-$326}

PKS B1257--326 is a flat spectrum, radio-loud quasar at $z = 1.256$ that 
exhibits IHV at frequencies of several GHz \citep{Bignall2003, Bignall2006}.  
Figure \ref{fig:1257} (left) shows the ISS time scales for PKS B1257--326 
observed on different days in 2000--2003 at 4.8 GHz by \citet{Bignall2006}
(kindly provided by Dr. Hayley Bignall).
The contours in the right panel show the range in transverse 
velocities that lead to fits with reduced $\chi^2 < 1.4$. With 26 observations 
and fitting for three parameters, the  good fits correspond to 
$\chi^2 \lesssim 1.2$.
In their analysis of the same time scale data but also including the ISS 
time delays between two widely separated radio 
telescopes, \citet{Bignall2006} found a best combined fit at velocities
$V_{iss,\alpha}=-49.2$~\kms\ and $V_{iss,\delta}=11.5$~\kms,
where the axial ratio was constrained to be less than 12.0.
They concluded that the scintillation pattern is highly elongated 
with an axial ratio $A \ge 12$, and the scattering screen is 
located within 10~pc of the Earth. We also find large axial ratios
for the ISS pattern.

Also plotted in the figure are the transverse velocities of the five nearby 
clouds located close to the line of sight to the quasar.
The transverse velocities for two clouds (Aur and Gem) give very good fits,
and the Gem cloud has a velocity very close to the best unconstrained 
five-parameter fit,
$V_{iss,\alpha}=-34$~\kms\ and $V_{iss,\delta}=0$~\kms. 
The ISS parameters for this fit are $A=4.1, \theta_{A}= 151.3^{\circ}$, with
values that vary systematically along the band of good fits in velocity space.
Note also that there is a small isolated region centered on
$V_{iss,\alpha}=-12$~\kms\ and $V_{iss,\delta}=-2$~\kms\ with
nearly the same minimum $\chi^2$ that is far from any cloud velocities.

The diamond symbol indicates the transverse velocity obtained by 
\citet{Bignall2006}, which is well outside the error bounds for 
the cloud velocities.  
They point out that the large anisotropy of the scintillation pattern
permits a range of best fit velocities that lie along a straight
line in the ($V_{iss,\alpha}, V_{iss,\delta}$) plane.
This line is situated along the best fit line in Figure~3(right).
Their best fit value of $V_{iss,\alpha}=-49.2$ km~s$^{-1}$ and 
$V_{iss,\delta}=11.5$ km~s$^{-1}$ is therefore poorly constrained.
However, it is remarkable that four 
of the five clouds have transverse velocities in the
band of velocities that provide ``reasonable'' fits to the ISS data. 
We show the time scale fit for the closest-fitting Gem cloud by the dashed
line in the left panel of the figure.
The clouds in this group are located within $\pm 10$~\kms\ of each other, 
suggesting a real physical association between these clouds and the plasma
producing the scintillation. 

Although the best fits (lowest $\chi^2$) in Figure~\ref{fig:1257} are for both 
the Aur and Gem clouds, the line of sight to the quasar passes through
the edge of the Gem cloud
but is about $15^{\circ}$ outside of the Aur cloud (Figure~\ref{fig:fig1}).
For this reason, we believe that
the scintillation screen lies at the edge of the Gem cloud, which lies within
6.7~pc of the Sun \citep{Redfield2007}. The orientation of the edge of the 
Gem cloud in celestial coordinates
is about $180^{\circ}$ compared to $\theta_A = 151.3^{\circ}$.

\subsection{\it Quasar B1519$-$273}

The 5 and 8~GHz time scale data from \citet{Jauncey2003} in the left panel of 
Figure \ref{fig:1519} are overplotted 
as in their Figure 1.  These data are not as extensive as for
the other two sources, and there are no errors given for the time scales.
Since the data do not show a consistent 
difference in time scale between the two frequencies,
we fit the two data sets simultaneously. 
The right panel (same format as for B1257--326) shows the contours that
define a good fit and the error ellipses for the four clouds near that line 
of sight.  Both the G and Leo clouds give good fits, lying within the 
$\chi^2 = 1.2$ contour.  We note that in the absence of error bars
for the time scales, the estimate of $\chi^2$ is based on the goodness of fit,
so that its normalization is not reliable.
The time scale model for the G cloud is overplotted in the 
left panel together with the best five parameter fit --
$V_{iss,\alpha}=-1.7$~\kms, $V_{iss,\delta}=-3.4$~\kms.
The ISS parameters for this fit are $A=3.5, \theta_{A}= 168^{\circ}$.
We note that these velocities 
are distinct from that of the Local Standard of Rest used by 
\citet{Jauncey2003} in their analysis of the same time scale data.
We conclude that the scintillation screen lies near the edge of the G cloud,
since the Leo cloud lies about $20^{\circ}$ from the source 
(Figure~\ref{fig:fig1}). The inner edge
of the G cloud lies closer than 1.3~pc from the Sun and may extend several 
parsecs away from the inner edge. The orientation of the G cloud edge 
in celestial coordinates is about
$110^{\circ}$ compared to $\theta_A = 168^{\circ}$.

\subsection{\it Quasar J1819+385}

Figure \ref{fig:1819} (left) shows the excellent 5~GHz ISS data  
overplotted from the years 
1999--2001 for the quasar J1819+385 observed by \citet{Dennett2003}.
This source shows the largest intraday variability presently known at 
radio wavelengths \citep{Dennett2001}. The best five parameter fit to 
these data is for $V_{iss,\alpha}=-1$~\kms\ and 
$V_{iss,\delta}=+20$~\kms, which is within the uncertainty of the Mic
cloud velocity. 

The right panel shows the narrow band in transverse velocity
space corresponding to fits to the ISS data with reduced $\chi^2 <1.2$. The 
transverse velocities of the Mic and LIC clouds are the best fits to the ISS 
data, although the G and Oph clouds provide fits with reduced 
$\chi^2 \approx 1.8$. Again, four of the six clouds have transverse 
velocities that give good-to-reasonable fits to the ISS data. 
The transverse velocity of the local
standard of rest is an unacceptable fit to the data, as already reported
by \citet{Dennett2003}. 

Our analysis, which follows that of \citet{Dennett2003}, agrees with their
result that the ISS data lie in a narrow region of  velocity space,
as shown in Figure \ref{fig:1819} (right).  By combining their time scale data
with time delay data on an intercontinental baseline, they arrived at a best
fit for $V_{iss,\alpha}=-33.5$~\kms\ and $V_{iss,\delta}=+13.5$~\kms\ 
relative to the local standard of rest.  This is plotted with 
the diamond symbol
in the heliocentric reference system that we use in this paper.  Their result
is well removed from the values for the local clouds, but as stated in a 
footnote in their paper, \citet{Dennett2003} used an algorithm 
for interpreting the
time delay data that was not corrected for anisotropic scintillation. We
therefore make no further use of that value plotted by the diamond.
Their Figures 10 and 11 give results for fitting to the annual changes
in time scale that yield an axial ratio $A>6$ oriented at 
$\theta_A = 7^{\circ}\pm 4^{\circ}$ This is consistent with our result, 
which gives a best fit at 
$A=11, \theta_A=5.5^{\circ}$.  They also conclude that the 
thin-scattering screen is located only 1--12 pc from the Sun.

The line of sight to the quasar lies well inside the LIC and close 
to the edge of the Mic cloud, 
the best fits in Figure~\ref{fig:1819}. It also lies close to the edge of the 
G cloud but some $5^{\circ}$ outside of the Oph cloud. We believe that the
most likely candidate cloud for the scintillation source is the Mic cloud since
\citet{Redfield2007} show that this cloud appears to be compressed by the
LIC and G clouds, and the Mic cloud has the highest temperature 
($9,900\pm 2,000$~K) of the 15 clouds they studied. The Mic cloud lies within 
5.1~pc of the Sun. The orientation of the Mic cloud edge 
in celestial coordinates is about $30^{\circ}$
compared to $\theta_A = 5.5^{\circ}$.

\section{CONCLUSIONS}

We have fit the ISS time scale observations of the three quasars that
show the largest amplitude intraday variations and have high quality data
sets.  For each quasar, our search of a large range of possible
transverse velocities of the scintillation screen yielded good
agreement with the annular variation of the ISS data for a relatively
narrow range of transverse velocities. For each quasar there are at
least two warm interstellar clouds that have transverse velocities
consistent with the acceptable range (reduced $\chi^2 \leq 1.2$),
but we propose that the scattering screen is located in one of the 
clouds near its edge. The transverse
velocity of the local standard of rest, the usual assumption in
previous studies, is an unacceptable solution for all three quasars.

In all cases, the scintillation screen is asymmetrical.
We compared the ellipse orientations $\theta_A$ with the
orientations of the nearest cloud edge to see whether the turbulence structure 
is elongated parallel to the cloud edge, perhaps by a magnetic field 
confining the outside of the cloud. We found no preferred alignments 
as the ellipse orientations differ typically by $25^{\circ}$ to
$50^{\circ}$ from the orientation of the edge of the cloud that we
believe is the location of the scattering screen.
This apparent misalignment may indicate that the cloud edges have complex 
geometries that our coarse fitting to a modest number of lines of sight
cannot reveal. Also \citet{Rickett2007} and \citet{Ramachandran2006}
show that the turbulent regions producing the observed scintillation are 
likely very small on the basis of the arrival times of pulses from the 
millisecond pulsar B1937+21 and other arguments.

Previous authors have placed the locations of the scattering screens 
near to the Sun on the basis of the scintillation time scales and the time 
delays between the scintillation flux variations as seen by two 
widely separated radio telescopes. Our analysis confirms and places more 
stringent constraints on the distances to the scattering screens. For example,
\citet{Bignall2006} place the scattering screen toward B1257--326 within 10~pc
of the Sun, and we argue that it is located within 6.7~pc. \citet{Dennett2003}
argue that the scattering screen for J1819+385 is located between 
1 and 12~pc from the Sun, and \citet{Macquart2007} present two models
that fit their data with the scattering screen located at $3.8\pm 0.3$~pc
or $2.0\pm 0.3$~pc. We argue that the screen toward this source is located 
at the edge of the Mic cloud within 5.1~pc of the Sun. If the Mic cloud 
is indeed compressed and heated 
by the LIC and G clouds, then the scattering screen 
in the Mic cloud is located much closer to the Sun than 5.1~pc.
The scattering screen toward B1519--273 near the edge of the G cloud
is likely closer than 3~pc from the Sun.

\begin{figure}
\epsscale{1.0}
\plotone{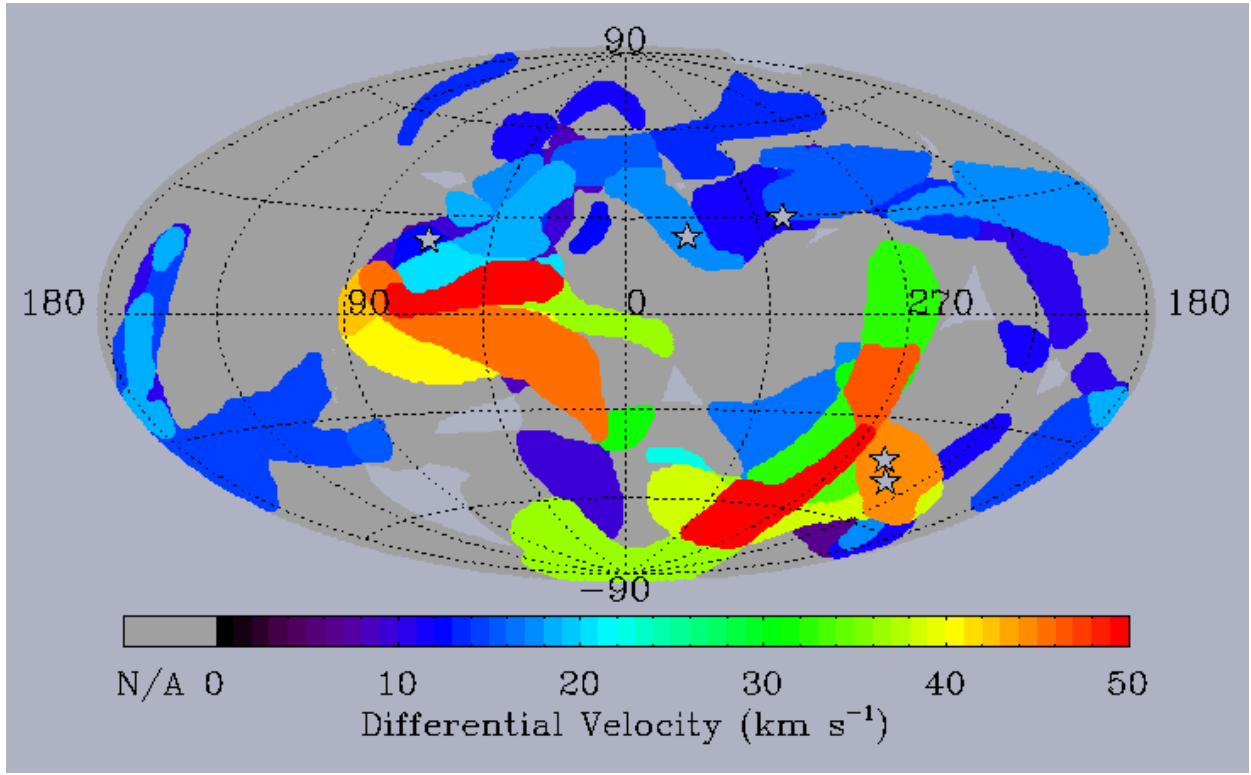}
\caption{Zones of LISM cloud possible interaction.  Colored regions
indicate directions in which multiple LISM clouds are detected along
the line of sight.  The white regions indicate where no LISM cloud is
detected and grey where only a single cloud is detected.  The color
coding signifies the maximum magnitude of the differential velocity
between clouds along the same line of sight.  Five scintillation sight
lines are shown here by star symbols.  The three northern hemisphere
sight lines (PKS B1257-326, QSO J1819+385, and QSO B1519-273) are
discussed in detail in Section~\ref{sec:fits}, while the two southern
hemisphere targets (PKS 0405-385 and PSR J0437-4715) await more data.
Note that all five sight lines are located in regions where multiple
clouds are located and where the differential velocity can be quite
high ($>$10 km~s$^{-1}$, and in two cases $>$45 km~s$^{-1}$).  If the
clouds are in contact with each other, such high differential
velocities are likely to induce shear flows and increase local
turbulence.\label{fig:fig2}}
\end{figure}

\begin{figure}
\epsscale{1.25}
\plottwo{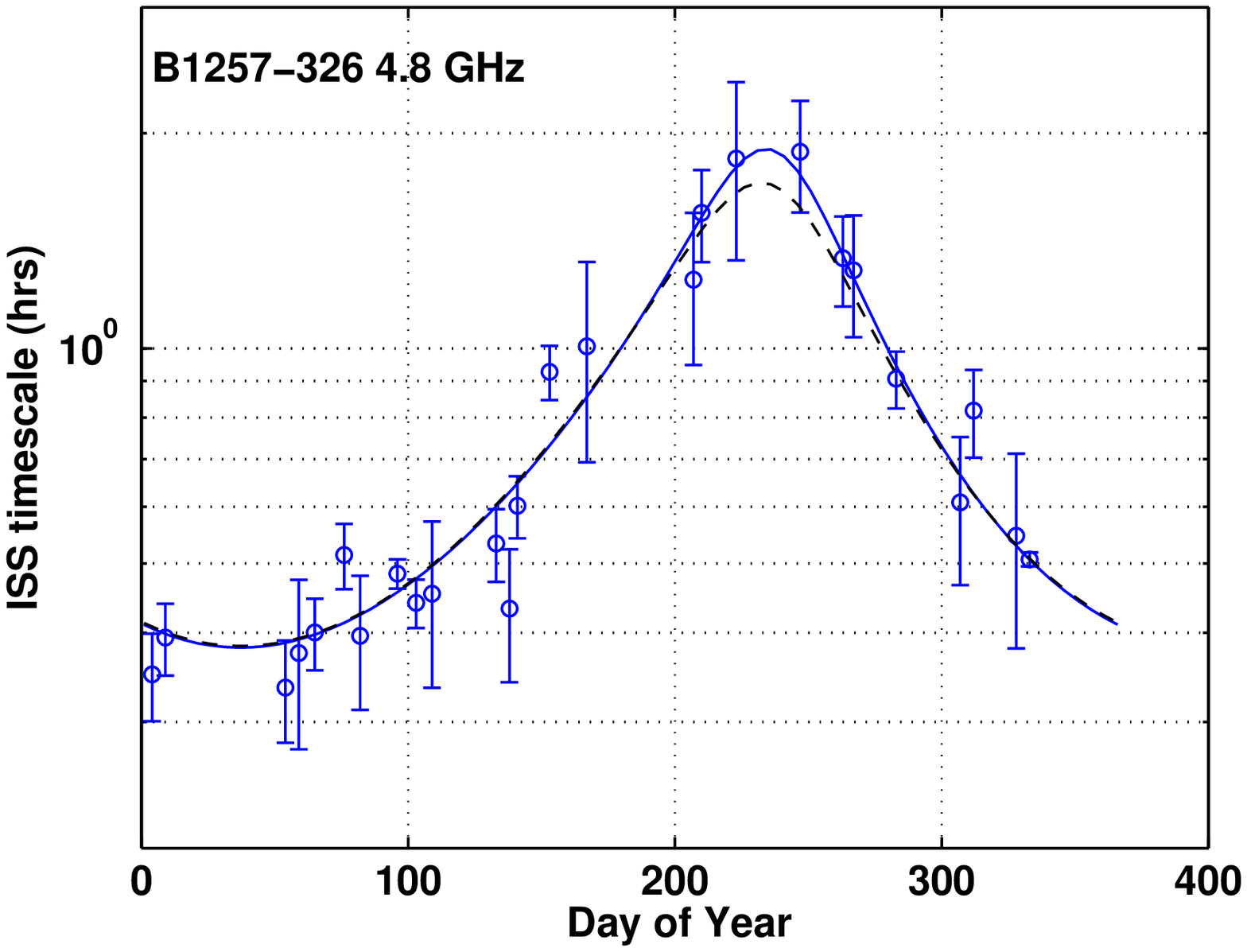}{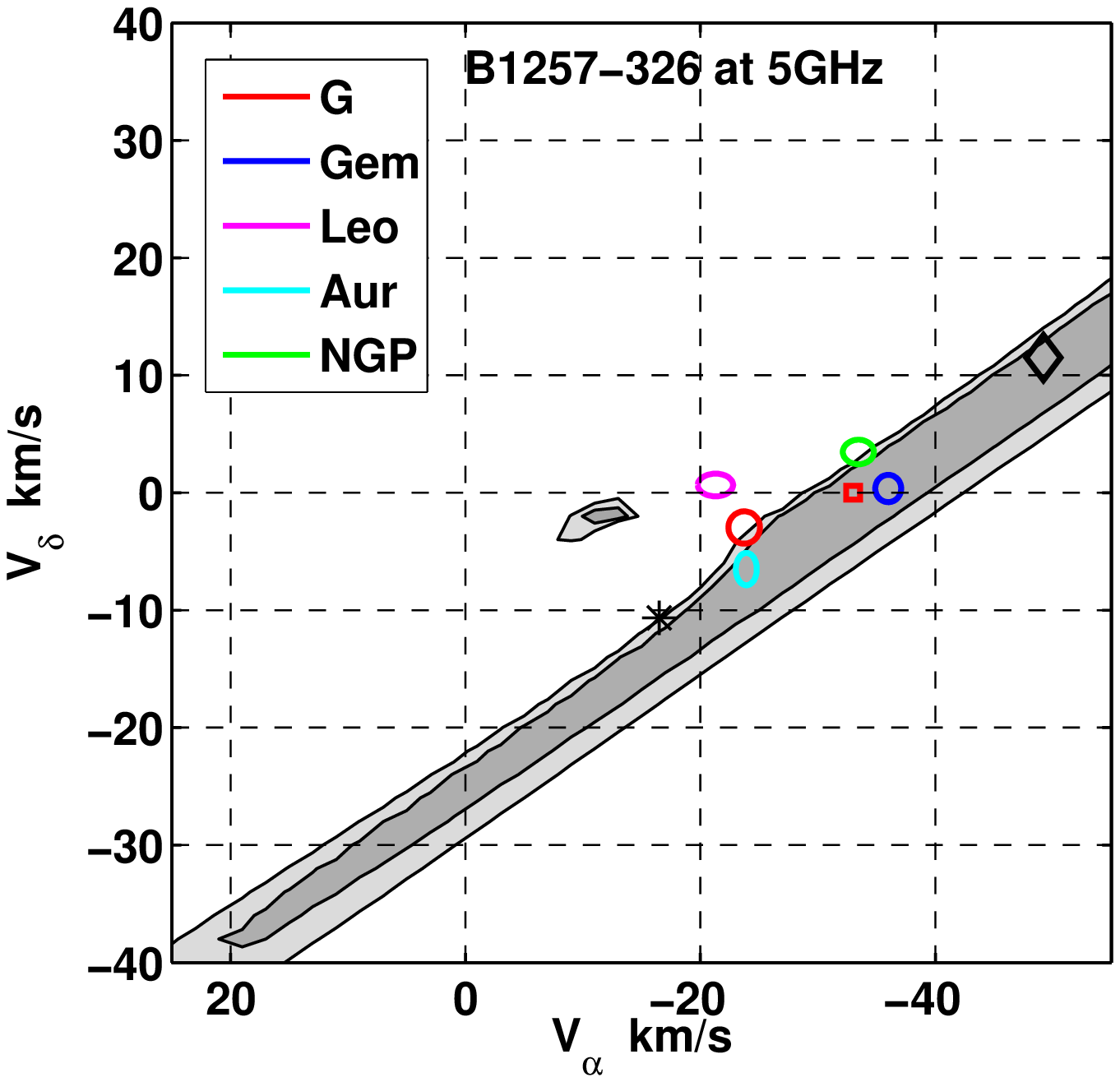}
\caption{{\em left}:
Intrahour 4.8 GHz flux variability time scale for PKS B1257--326 
from \citet{Bignall2006}. The solid line is the best 
unconstrained five-parameter fit, $V_{iss,\alpha}=-34$~\kms\ and
$V_{iss,\delta}=+0$~\kms.  The dashed line is for a 3-parameter fit assuming
the transverse velocity for the Gem cloud.
{\em right}: 
Velocity plot for PKS B1257--326. The entire plotted region of
transverse velocity space ($V_{iss,\alpha}$,$V_{iss,\delta}$)
was searched for the best three parameter fits to 
the ISS time scale data. The contours show the (reduced) $\chi^2$ 
(at 1.2 \& 1.4) for 
three-parameter fits to the data. The error ellipses are overplotted
for the transverse velocities of clouds along that line of sight.
Good fits (inner contour at $\chi^2 < 1.2$) are obtained for 
the Gem and Aur clouds. The black star is the projected transverse 
velocity of the local standard of rest. The best five parameter fit 
(square symbol) is very close to the transverse velocity of the Gem cloud.  
The diamond symbol is the best combined fit to the same time scale data
and the ISS time delays by \citet{Bignall2006}.
\label{fig:1257}}
\end{figure}

\begin{figure}
\epsscale{1.25}
\plottwo{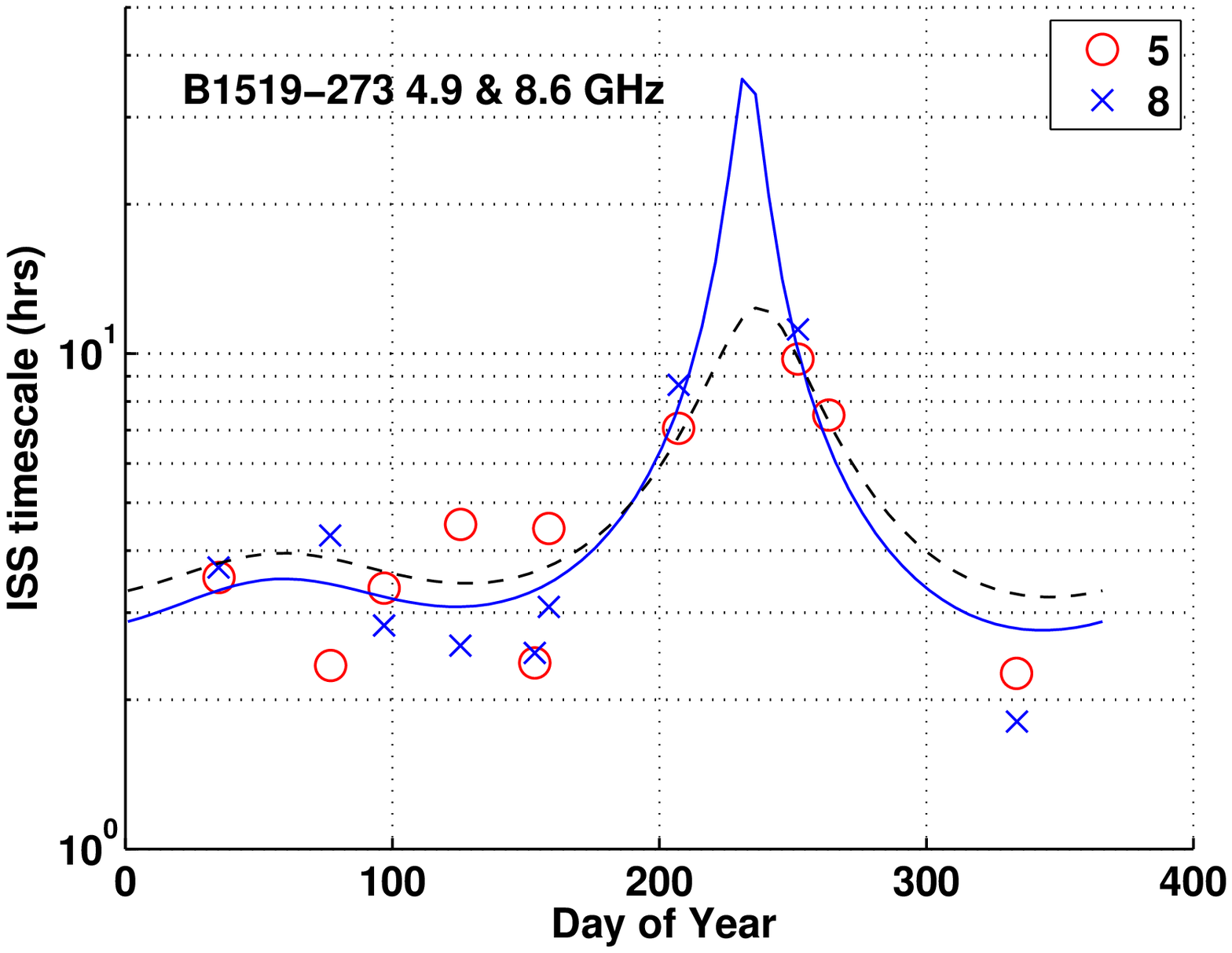}{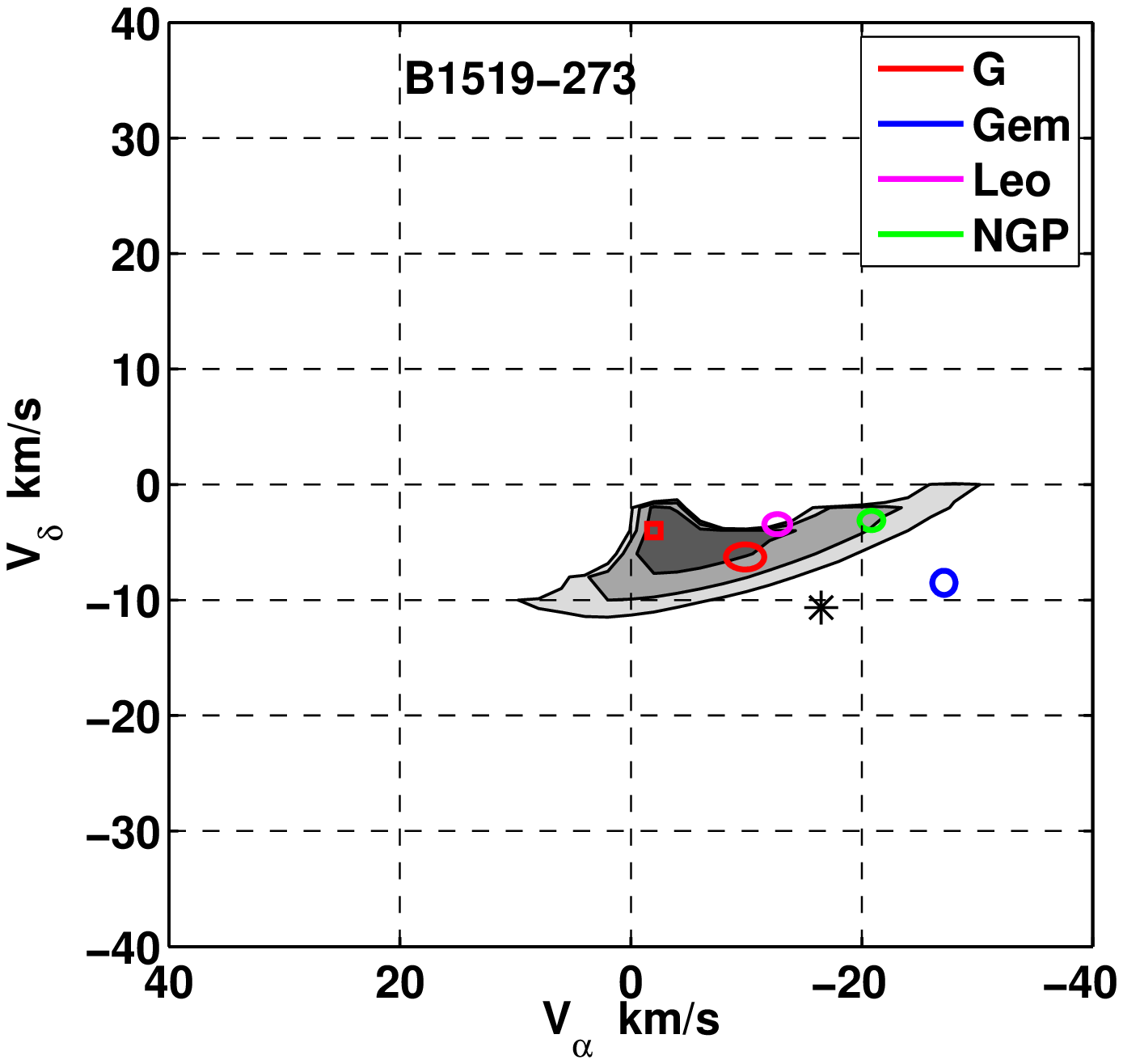}
\caption{{\em left}:
Intrahour 4.8 GHz flux (circles) and 8 GHz ($\times$ symbols) variability 
time scale data for quasar B1519--273. The solid line is the best 
unconstrained five-parameter fit, $V_{iss,\alpha}=+2$~\kms\ and
$V_{iss,\delta}=-3$~\kms\ ($\chi^2=0.66$).  The dashed 
line is for a 3-parameter fit assuming
the transverse velocity for the G cloud.
{\em right}: 
Velocity plot for B1519--273. The entire plotted region of
transverse velocity space ($V_{iss,\alpha}$,$V_{iss,\delta}$)
was searched for the best three-parameter fits to 
the ISS time scale data. The grey scale shows the (reduced) $\chi^2$ for 
three-parameter fits to the data. The best fits ($\chi^2 < 1.2$, darkest grey) 
are for the G and Leo clouds. The projected transverse velocity of the 
local standard of rest (asterisk symbol) and the
best five-parameter fit (square symbol) are not close to the transverse 
velocity of any cloud.  
\label{fig:1519}}
\end{figure}

\begin{figure}
\epsscale{1.25}
\plottwo{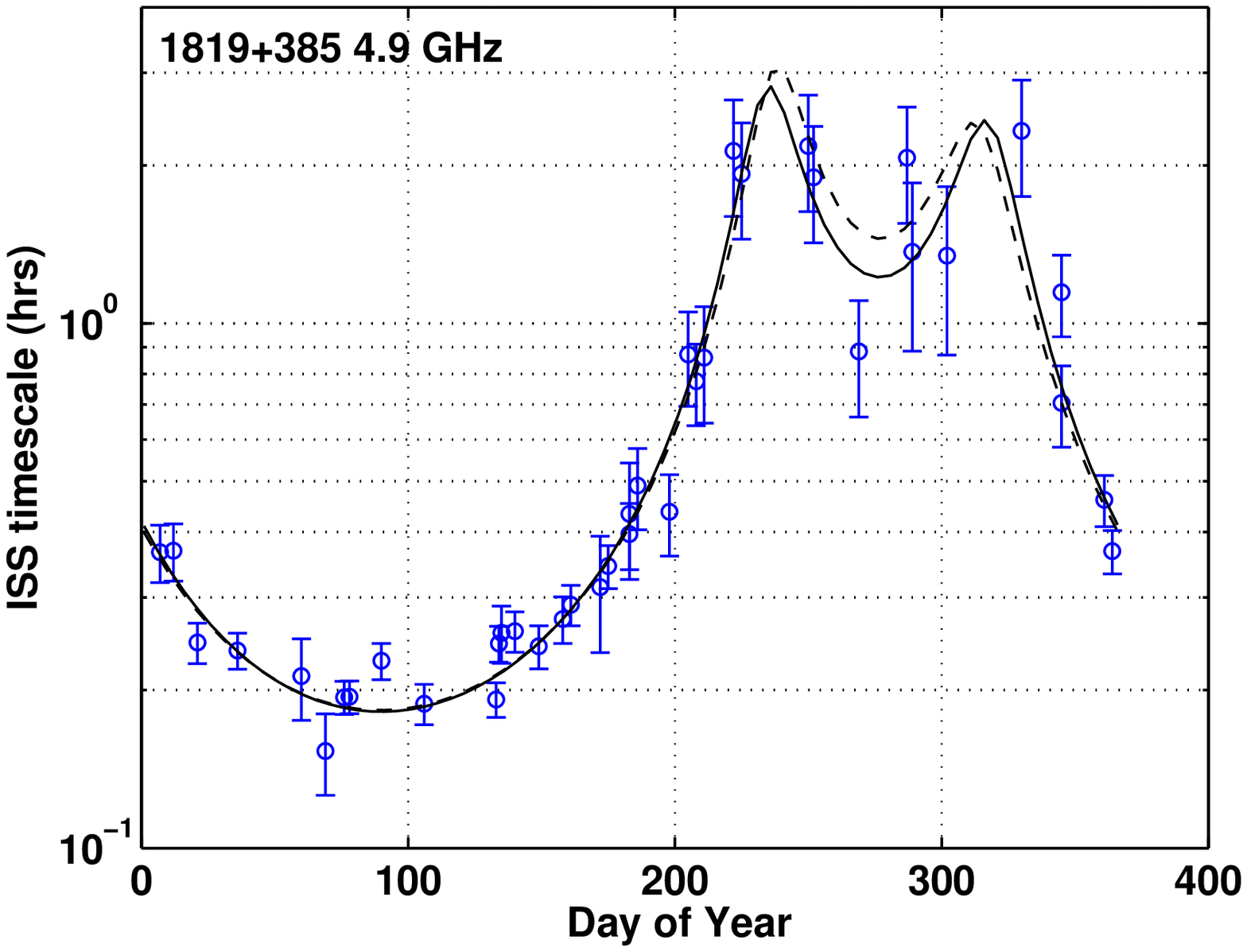}{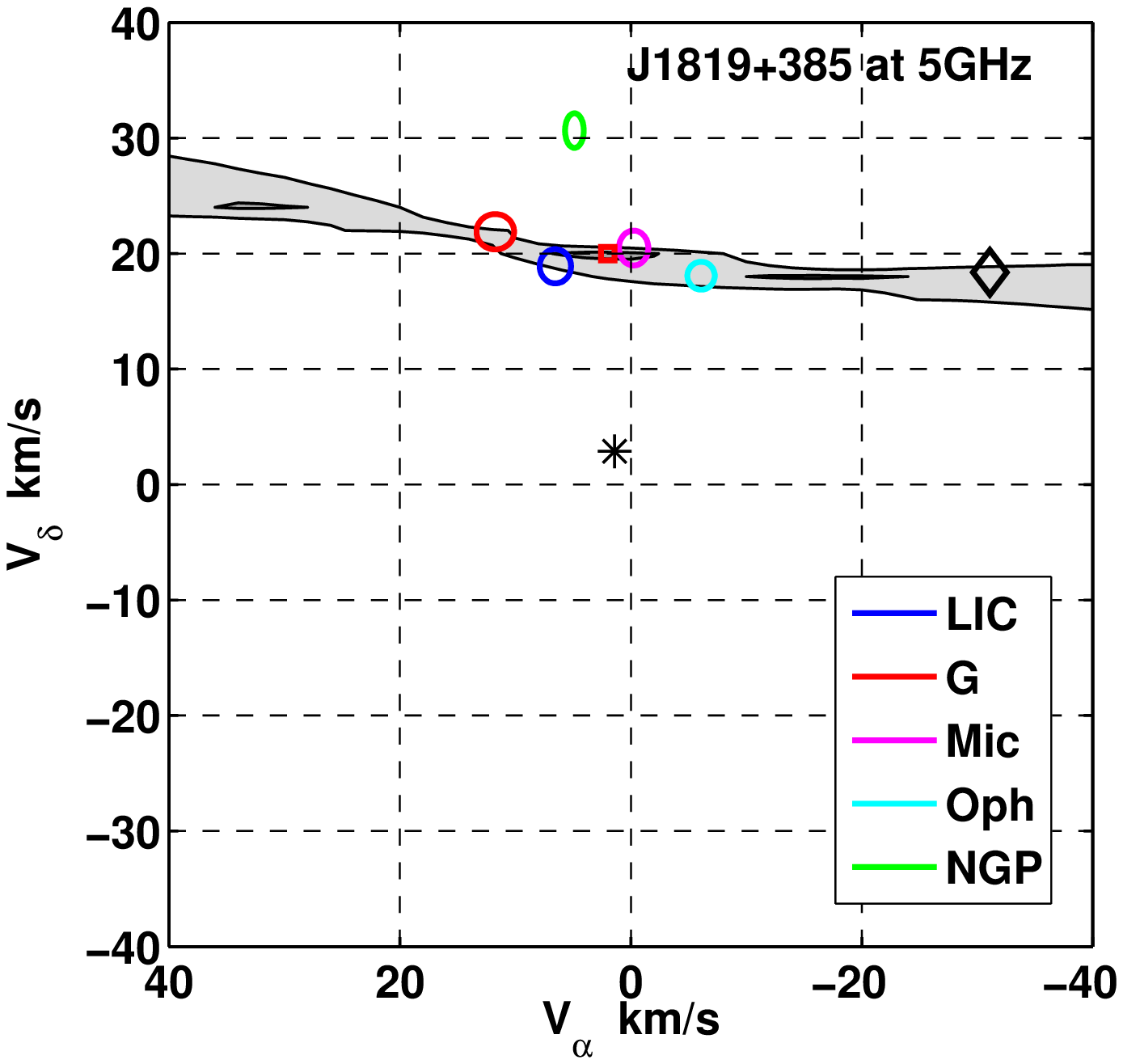}
\caption{{\em left}:
Intrahour 4.9 GHz flux variability time scale data for quasar J1819+385
\citep{Dennett2003}. The solid line is the best 
unconstrained five-parameter fit, $V_{iss,\alpha}=+1$~\kms\ and
$V_{iss,\delta}=+20$~\kms.  The dashed line is for a three-parameter fit 
assuming the transverse velocity for the Mic cloud.
{\em right}: 
Velocity plot for J1819+385. The entire plotted region of
transverse velocity space ($V_{iss,\alpha}$,$V_{iss,\delta}$)
was searched for the best three-parameter fits to 
the ISS time scale data. The grey scale shows the (reduced) $\chi^2$ for 
three-parameter fits to the data. The best fits ($\chi^2 < 1.2$) are for 
the Mic and LIC clouds, although the G and Oph clouds provide fits with
reduced $\chi^2 \approx 1.8$. 
The  projected transverse velocity of the local 
standard of rest is indicated by the asterisk symbol.
The best five-parameter fit (square symbol) is very close to the transverse 
velocity of the Mic cloud.  The diamond symbol indicates the transverse 
velocity determined by \citet{Dennett2003}.
\label{fig:1819}}
\end{figure}

An important result is that for each source, the line of sight to the 
quasar passes through or very near to at least two warm clouds. Although we
know only upper limits to the distances to most of these clouds, 
they are all nearby and could interact. For example, 
interaction of the LIC and G clouds is likely responsible for the shape and 
high temperature of the Mic cloud. Since typical velocity differences between
clouds along the three lines of sight are 10--20~\kms, the interactions could
easily produce turbulence that would be responsible for the scintillation. 
Also, cloud edges are regions of high ionization because of the absence of 
shielding from the strong UV radiation from hot stars and white dwarfs.
Figure~\ref{fig:fig2} shows areas on the sky where the velocity differences
between clouds along the line of sight are large and where cloud-cloud 
interactions could produce turbulence. We predict that these are regions where
large amplitude IDV sources will be discovered. Two examples that deserve 
further study are the two IDV sources located 
in the southern hemisphere (PKS 0405--385 and PSR J0437--4715) that are 
plotted in Figure~\ref{fig:fig2} but are not analyzed 
in this paper as high quality ISS data are not yet published.

\acknowledgments

We thank Dr. Hayley Bignall for providing the data on PKS B1257-326
that we have analyzed in this paper. We also than Dr. Jean-Pierre
Macquart for discussions on this topic and the referee for his very
useful suggestions.  S.R. would like to
acknowledge support provided by NASA through Hubble Fellowship grant
HST-HF-01190.01 awarded by the Space Telescope Science Institute,
which is operated by the Association of Universities for Research in
Astronomy, Inc., for NASA, under contract NAS 5-26555.  

\clearpage

\clearpage

\clearpage

\clearpage

\clearpage

\clearpage

\begin{deluxetable}{lcccccc}
\tablewidth{0pt}
\tabletypesize{\footnotesize}
\tablecaption{Warm LISM Cloud Kinematics Along Radio Scintillation Sight 
Lines\label{tab:lismvecs}}
\tablehead{Sight Line & $l$ & $b$ & Nearby & $v_{\rm radial}$ & $v_{\rm RA}$ & 
$v_{\rm Dec}$ \\ 
 & (deg) & (deg) & Cloud & (km s$^{-1}$) & (km s$^{-1}$) & (km s$^{-1}$)}
\startdata
PKS B1257-326  & 305.2 &  29.9 & NGP  & --15.42 $\pm$ 1.09 & 
--33.45 $\pm$ 1.32 & 3.47 $\pm$ 0.87 \\
               &       &       & Gem  & --4.75 $\pm$ 0.97 & 
--35.99 $\pm$ 1.10 & 0.36 $\pm$ 0.94 \\
               &       &       & Leo  & --9.96 $\pm$ 1.17 & 
--21.28 $\pm$ 1.50 & 0.66 $\pm$ 0.90 \\
               &       &       & G    & --17.45 $\pm$ 1.11 & 
--23.73 $\pm$ 1.34 & --2.96 $\pm$ 1.26 \\
               &       &       & Aur  & --4.72 $\pm$ 1.16 & 
--23.91 $\pm$ 0.89 & --6.50 $\pm$ 1.26 \\
QSO J1819+385  & 66.2  & 22.5  & LIC  & --13.05 $\pm$ 1.26 & 6.54 $\pm$ 1.36 & 
18.90 $\pm$ 1.47 \\
               &       &       & Mic  & --19.74 $\pm$ 1.31 & 
--0.24 $\pm$ 1.27 & 20.48 $\pm$ 1.51 \\
               &       &       & Oph  & --26.01 $\pm$ 0.95 & 
--6.08 $\pm$ 1.18 & 18.08 $\pm$ 1.23 \\
               &       &       & G    & --16.10 $\pm$ 1.03 & 
11.74 $\pm$ 1.63 & 21.89 $\pm$ 1.50 \\
               &       &       & Aql  & --34.85 $\pm$ 1.10 & 
42.70 $\pm$ 1.17 & 19.90 $\pm$ 0.96 \\
               &       &       & NGP  & --20.14 $\pm$ 1.16 & 4.89 $\pm$ 0.81 & 
30.65 $\pm$ 1.49 \\
QSO B1519-273  & 339.6 & 24.4  & G    & --27.19 $\pm$ 1.08 & 
--9.89 $\pm$ 1.66 & --6.27 $\pm$ 1.07 \\
               &       &       & Gem  & --22.59 $\pm$ 1.02 & 
--27.11 $\pm$ 1.00 & --8.51 $\pm$ 0.93 \\
               &       &       & NGP  & --30.42 $\pm$ 1.28 & 
--20.83 $\pm$ 1.03 & --3.11 $\pm$ 0.78 \\
               &       &       & Leo  & --19.49 $\pm$ 1.45 & 
--12.68 $\pm$ 1.12 & --3.43 $\pm$ 0.85 \\
\enddata
\tablecomments{All velocities are heliocentric.}
\end{deluxetable}  

\end{document}